\documentclass[10pt,fleqn]{article}

\usepackage{amsmath,amssymb}
\usepackage{graphicx}
\usepackage[top=0.75in, bottom=0.75in, left=0.75in, right=0.75in, dvips]{geometry}

\def\tref#1#2#3{{#1} (#2) #3}
\newcommand{\ii}{\ensuremath{\mathrm{i}}}
\newcommand{\e}{\ensuremath{\mathrm{e}}}
\newcommand{\ppi}{\ensuremath{\mathrm{\pi}}}
\newcommand{\dif}[1]{\ensuremath{ \mathrm{d}\,#1 }}
\newcommand{\borel}{\ensuremath{\mathcal{B}}}
\newcommand{\il}{\ensuremath{\mathcal{L}^{-1}}}
\newcommand{\qcd}{\ensuremath{\mathrm{QCD}}}

\begin{document}
\title{
Gaussian Sum-Rule Analysis of Scalar Gluonium and Quark Mesons
}
\author{T.G.\ Steele\\
{\it Department of Physics \& Engineering
    Physics, University of Saskatchewan}\\ 
{\it Saskatoon, SK~~S7N 5E2, Canada}\\
D.\ Harnett\\
{\it Department of Physics, University
    College of the Fraser Valley}\\
{\it Abbotsford, BC~~V2S 7M8, Canada}\\
G.\ Orlandini\\
{\it Dipartimento di Fisica and INFN Gruppo Collegato di Trento}
\\
{\it Universit\`a di Trento, I-38050 Povo, Italy}}

\maketitle

\begin{abstract}
Gaussian sum-rules, which are related to a two-parameter Gaussian-weighted integral of a hadronic spectral function,
are able to examine the possibility that  more than one resonance makes a significant contribution to the spectral function.
The Gaussian sum-rules, including instanton effects, for scalar gluonic and non-strange scalar quark currents clearly indicate a distribution of the resonance strength in their respective spectral functions.
Furthermore, analysis of a two narrow resonance model leads to excellent agreement between theory and phenomenology
in both channels.
The scalar quark and gluonic sum-rules  are remarkably consistent in their prediction
of masses of approximately $ 1\;{\rm GeV}$ and  $1.4\;{\rm GeV}$  within this model.
Such a similarity would be expected from hadronic states which are mixtures of gluonium and quark mesons.
\end{abstract}

\section{Introduction}
The multitude of scalar hadronic states with masses above $1\;{\rm GeV}$~\cite{pdg} is frequently noted
as evidence that a $q\bar q$ nonet is  insufficient to accommodate these states, as would be expected
from the existence of a (scalar) gluonium state.
Gaussian QCD sum-rules have been shown to be sensitive to the hadronic spectral function over a broad energy range,
and analysis techniques have been developed to exploit this dependence to determine how resonance strength is
distributed in the spectral function~\cite{orl00,har01,utica02}.  Thus Gaussian sum-rules  provide a valuable
technique for  determining how the quark and gluonium
content is distributed amongst scalar hadronic states.

The simplest Gaussian sum-rule (GSR) has the form~\cite{gauss}
\begin{equation}
G_0\left(\hat s,\tau\right)=\frac{1}{\sqrt{4\ppi\tau}} \int\limits_{t_0}^\infty
\exp\left[\frac{-\left(t-\hat{s}\right)^2}{4\tau}\right]\,\frac{1}{\ppi}\rho(t)\;\dif{t}
\quad,\quad\tau>0
\label{basic_gauss}
\end{equation}
and relates the QCD prediction $G_0\left(\hat s,\tau\right)$ to an integral of its associated hadronic spectral function
$\rho(t)$.  The smearing of the spectral function by the Gaussian kernel peaked at $t=\hat s$ through the
(approximate) region $\hat s-2\sqrt{\tau}\le t\le \hat s+2\sqrt{\tau}$
provides a clear conceptual implementation
of quark-hadron duality.  The width of this duality interval is constrained by QCD since renormalization-group improvement
of the QCD (left-hand) side of~\eqref{basic_gauss} results in identifying  the renormalization scale  $\nu$ through
$\nu^2=\sqrt{\tau}$~\cite{orl00,gauss}; therefore it is not possible to achieve the formal $\tau\to 0$ limit
where complete knowledge of the spectral function could be obtained through
\begin{equation}
\lim_{\tau\to 0}G_0\left(\hat s,\tau\right)=\frac{1}{\ppi}\rho\left(\hat s\right)\quad ,\quad \hat s>t_0 \quad .
\label{gauss_limit}
\end{equation}
The variable $\hat{s}$ in~\eqref{basic_gauss}, on the other hand, is unconstrained by QCD, and so the $\hat{s}$
dependence of $G_0\left(\hat s,\tau\right)$ can be used to probe the behaviour of the smeared spectral function, and hence
the essential features of $\rho(t)$.

An interesting feature of the GSR~\eqref{basic_gauss} is its ability to study excited and ground states  with similar
sensitivity.
For example, as $\hat s$ passes through $t$ values corresponding to resonance peaks, the Gaussian
kernel reaches its maximum value.
Thus any features of the spectral function strong enough to be isolated from the
continuum will be revealed through the GSR.
In this regard, GSRs should be contrasted with Laplace sum-rules
 \begin{equation}
R\left(\Delta^2\right)=\frac{1}{\pi}\int\limits_{t_0}^\infty\exp{\left(-\frac{t}{\Delta^2}\right)}\rho(t)\, \dif{t}
\quad ,
\label{basic_laplace}
\end{equation}
which exponentially suppress excited states in comparison to the ground state.

In what follows, the original formulation of GSRs~\cite{gauss} will be reviewed along with corresponding GSR
analysis techniques~\cite{orl00,har01}.
A GSR moments analysis for scalar gluonic and quark correlation
functions will be presented in order to demonstrate that the associated spectral functions have
a distribution of resonance strength.
In addition, mass and relative coupling predictions specific to a double narrow resonance model will be determined~\cite{orl00,har01}.
The major emphasis of this paper is to highlight the remarkable consistency
between GSR mass predictions extracted from the scalar gluonic channel and from the scalar-isoscalar quark channel,
a result indicative of the existence of hadronic states which are mixtures of quark mesons and
gluonium.

\section{Formulation and Analysis Techniques for Gaussian Sum-Rules}
\begin{figure}[htb]
\centering
\includegraphics[scale=0.4]{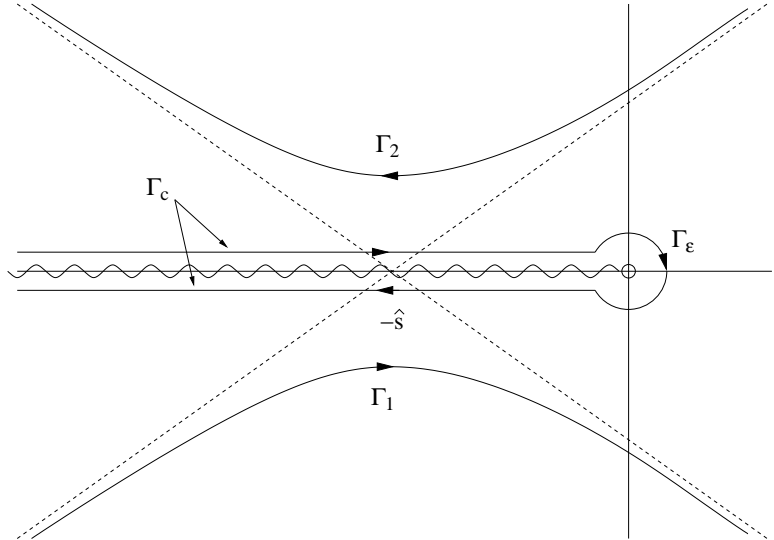}
\caption{{\small Contours of integration $\Gamma_1+\Gamma_2$ and $\Gamma_c+\Gamma_\epsilon$ defining the
Gaussian sum-rule (see~\eqref{finish} and~\eqref{finishDerek} respectively).
The wavy line on the negative real axis denotes the branch cut of $\Pi(w)$.
}}
\label{cont_fig}
\end{figure}

Gaussian sum-rules are based on QCD correlation functions of renormalization-group invariant
composite operators $J(x)$
\begin{equation}
  \Pi\left(Q^2\right)=\ii\int\mathrm{d}^4x\;
  \e^{\ii q\cdot x}\left\langle O \vert T\left[ J(x) J(0)\right] \vert O \right\rangle
  \quad,\quad Q^2\equiv -q^2
\label{basic_corr_fn}
\end{equation}
which, in turn, satisfy dispersion relations appropriate to the asymptotic form
of the correlator in question.
For example, the scalar gluonic correlation function $\Pi\left(Q^2\right)$ (see~\eqref{glue_corr} and~\eqref{glue_curr}) satisfies the following dispersion relation with three subtraction constants
\begin{equation}
  \Pi\left(Q^2\right)-\Pi(0)-Q^2\Pi'(0)-\frac{1}{2}Q^4\Pi''(0)
  =-\frac{Q^6}{\pi}\int\limits_{t_0}^\infty
  \frac{\rho(t)}{t^3\left(t+Q^2\right)}\;\dif{t}
\quad.
\label{disp_rel}
\end{equation}
In general, the quantity $\rho(t)$ is the spectral function appropriate
to the quantum numbers of the given current, and it should be noted that in certain situations (such as the scalar
gluonic correlation function) the subtraction constant $\Pi(0)$ is determined by a low-energy theorem \cite{let} (see~\eqref{letDerek}).
Undetermined dispersion-relation constants and field-theoretical divergences are eliminated
through the GSR~\cite{har01}\footnote{This definition is a natural generalization of that given in~\cite{gauss}.
To recover the original GSR, we simply let $k=0$ in~(\ref{srdef}).}
\begin{equation}\label{srdef}
   G_k(\hat{s},\tau)\equiv \sqrt{\frac{\tau}{\ppi}}\borel
   \left\{ \frac{(\hat{s}+\ii\Delta)^k \Pi(-\hat{s}-\ii\Delta)
         - (\hat{s}-\ii\Delta)^k \Pi(-\hat{s}+\ii\Delta) }{\ii\Delta}
   \right\}
\end{equation}
where $k= -1,0,1, \ldots$ and  where the Borel transform $\borel$ is defined by
\begin{equation}\label{borel}
  \borel\equiv \lim_{\stackrel{N,\Delta^2\rightarrow\infty}{\Delta^2/N\equiv 4\tau}}
  \frac{(-\Delta^2)^N}{\Gamma(N)}\left( \frac{\mathrm{d}}{\mathrm{d}\Delta^2}\right)^N \quad.
\end{equation}
The dispersion relation~\eqref{disp_rel} in conjunction with definition~\eqref{srdef} together
yield the following family of GSRs
\begin{equation}
 G_{k}(\hat{s},\tau)+ \delta_{k\,-1}\frac{1}{\sqrt{4\pi\tau}}
 \exp\left( \frac{-\hat{s}^2}{4\tau}\right)\Pi(0)
 = \frac{1}{\sqrt{4\pi\tau}}\int_{t_0}^{\infty} t^k
   \exp\left[ \frac{-(\hat{s}-t)^2}{4\tau}\right]\frac{1}{\pi}\rho(t)\;\dif{t}
\label{gauss_family}
\end{equation}
where the identity
\begin{equation}
  \borel\left[\frac{1}{\Delta^2+a}\right]
  =\frac{1}{4\tau}\exp\left(\frac{-a}{4\tau}\right)
  \quad,\quad
  n=0,1,2,\ldots
\label{borelDerek}
\end{equation}
has been used to simplify the phenomenological (right-hand) side of~\eqref{gauss_family} as well as the term on the QCD
side proportional to $\Pi(0)$.
Evidently the $k=-1$ sum-rule can only be defined in cases where there exists an appropriate low-energy theorem.
Calculation of $G_k(\hat{s},\tau)$ is achieved through an identity relating~(\ref{borel}) to the inverse
Laplace transform~\cite{gauss}
\begin{equation}\label{bor_to_lap}
  \borel [f(\Delta^2)] = \frac{1}{4\tau}\il [f(\Delta^2)]
\end{equation}
where, in our notation,
\begin{equation}
  \il [f(\Delta^2)] = \frac{1}{2\ppi \ii} \int\limits_{a-\ii\infty}^{a+\ii\infty}
    f(\Delta^2) \exp\left( \frac{\Delta^2}{4\tau} \right) \dif{\Delta^2}
\end{equation}
with $a$ chosen such that all singularities of $f$ lie to the left of $a$
in the complex $\Delta^2$-plane.
Through a simple change of variables, the calculation of the GSR reduces to~\cite{har01}
\begin{equation}
  G_k(\hat{s},\tau)  = \frac{1}{\sqrt{4\ppi\tau}}\frac{1}{2\ppi \ii}
  \int_{\Gamma_1 +\Gamma_2} (-w)^k  \exp
  \left[ \frac{-(\hat{s}+w)^2}{4\tau}\right] \Pi(w)\;\dif{w}
  \label{finish}
\end{equation}
which, by suitably deforming the complex contours $\Gamma_1$ and $\Gamma_2$, can
be recast in the form
\begin{equation}
  G_k(\hat{s},\tau)  = -\frac{1}{\sqrt{4\ppi\tau}}\frac{1}{2\ppi \ii}
  \int_{\Gamma_c +\Gamma_{\epsilon}} (-w)^k \exp
  \left[ \frac{-(\hat{s}+w)^2}{4\tau}\right]\Pi(w)\;\dif{w}
\label{finishDerek}
\end{equation}
where the trajectories $\Gamma_1$, $\Gamma_2$, $\Gamma_c$, and $\Gamma_{\epsilon}$ are
all depicted in Figure~\ref{cont_fig}.  Further simplification of~\eqref{finishDerek} requires
a specific correlator $\Pi$ \textit{i.e.}~\eqref{glue_corr} or~\eqref{quark_corr}.

Next, we impose a fairly general resonance(s) plus continuum model
\begin{equation}
  \rho(t)=\rho^{{\rm had}}(t)+\theta\left(t-s_0\right){\rm Im}\Pi^{\qcd}(t)
\label{respcont}
\end{equation}
where $s_0$ represents the onset of the QCD continuum.
The continuum contribution of~\eqref{respcont} to the right-hand side of~\eqref{gauss_family} is
\begin{equation}\label{continuum}
   G_k^{{\rm cont}} (\hat{s},\tau,s_0) = \frac{1}{\sqrt{4\pi\tau}}   \int_{s_0}^{\infty} t^k
   \exp \left[ \frac{-(\hat{s}-t)^2}{4\tau} \right]  \frac{1}{\pi} {\rm Im} \Pi^{\qcd}(t)\; \dif{t}\quad ,
\end{equation}
and is combined with $G_k\left(\hat s,\tau\right)$ to obtain the total QCD contribution
\begin{equation}
  G_k^{\qcd}\left(\hat{s},\tau,s_0\right) \equiv G_k\left(\hat{s},\tau\right) -  G_k^{{\rm cont}} \left(\hat{s},\tau,s_0\right) \quad ,
\label{blah}
\end{equation}
resulting in the final relation between the QCD and hadronic sides of the GSRs
\begin{equation}
 G_{k}^\qcd\left(\hat{s},\tau,s_0\right)+ \delta_{k\,-1}\frac{1}{\sqrt{4\pi\tau}}
 \exp\left( \frac{-\hat{s}^2}{4\tau}\right)\Pi(0)
  =\frac{1}{\sqrt{4\pi\tau}}\int_{t_0}^{\infty} t^k
   \exp\left[ \frac{-(\hat{s}-t)^2}{4\tau}\right] \frac{1}{\pi} \rho^{\rm had}(t)\; \dif{t}
   \quad.
\label{final_gauss}
\end{equation}

Original studies involving the GSRs exploited the diffusion equation
\begin{equation}
\frac{\partial^2 G_0\left( \hat s,\tau\right)}{\partial\hat s^2}=
\frac{\partial G_0\left( \hat s,\tau\right)}{\partial \tau}
\label{diffusion}
\end{equation}
which follows from~\eqref{gauss_family} for $k=0$.
In particular, when $\rho^{{\rm had}}(t)$ (see~\eqref{respcont}) is evolved through
the diffusion equation~\eqref{diffusion}, it only reproduces the QCD prediction at large energies ($\tau$ large)
if the resonance and continuum contributions are balanced through the $k=0$ member of the finite-energy sum-rule family \cite{gauss}
\begin{equation}
F_k\left(s_0\right)=\frac{1}{\pi}\int\limits_{t_0}^{s_0} t^k\rho^{\rm had}(t)\;\dif{t}\quad .
\label{basic_fesr}
\end{equation}

An additional connection between the finite-energy sum-rules \eqref{basic_fesr} and the GSRs can be found by integrating both sides of~\eqref{final_gauss} with respect to $\hat{s}$ to obtain
\begin{equation}
  \int\limits_{-\infty}^\infty G_k^{\qcd}(\hat{s},\tau,s_0)\;\dif{\hat{s}}+\delta_{k\,-1}\Pi(0)
  =\frac{1}{\pi}\int\limits_{t_0}^{\infty} t^k\rho^{{\rm had}}(t)\;\dif{t}\quad ,
  \label{tom_norm_2}
\end{equation}
indicating that the finite-energy sum-rules \eqref{basic_fesr} are related to the normalization of the GSRs. Thus the information
independent of the finite-energy sum-rule constraint following from the diffusion equation analysis \cite{gauss} is contained in the
{\em normalized} Gaussian sum-rules (NGSRs) \cite{orl00}
\begin{gather}
  N^{\qcd}_k (\hat{s}, \tau, s_0) =
  \frac{G^{\qcd}_{k} (\hat s, \tau, s_0) + \delta_{k\,-1}\frac{1}{\sqrt{4\pi\tau}}
  \exp\left(\frac{-\hat{s}^2}{4\tau}\right)
  \Pi(0) }{M^{\qcd}_{k,0} (\tau, s_0)+\delta_{k\,-1}\Pi(0)}
\label{tom_norm_srk}
\\
  M_{k,n}(\tau, s_0)
 =\int\limits_{-\infty}^\infty \hat{s}^n G_k (\hat s,\tau, s_0)\;\dif{\hat{s}}
  \quad,\quad n=0,1,2,\ldots\quad,
\label{moments}
\end{gather}
which are related to the hadronic spectral function via
\begin{equation}\label{ngsr}
   N_k^{\qcd}(\hat{s},\tau,s_0) = \frac{ \frac{1}{\sqrt{4\pi\tau}} \int_{t_0}^{\infty} t^k
   \exp\left[\frac{-(\hat{s}-t)^2}{4\tau} \right] \rho^{{\rm had}}(t)\;\dif{t}}{\int_{t_0}^{\infty} t^k
   \rho^{{\rm had}}(t)\;\dif{t}} \quad.
\end{equation}

\section{Gaussian Sum-Rules for Scalar Gluonic and Non-Strange Scalar Quark Currents}
At leading order in the quark mass for $n_f$ flavours, scalar hadronic states can
either be probed through the correlation function of the
(renormalization-group invariant) gluonic current
\begin{gather}
  \Pi_g\left(Q^2\right)=\ii\int\,\mathrm{d}^4x\,\e^{\ii q\cdot x}\left\langle O
  \vert T\left[ J_g(x)J_g(0)\right] \vert O \right\rangle
  \quad,\quad Q^2\equiv -q^2
  \label{glue_corr}
\\
  J_g(x)=-\frac{\pi^2}{\alpha\beta_0}\beta\left(\alpha \right)G^a_{\mu\nu}(x)G^a_{\mu\nu}(x)
  \label{glue_curr}
\end{gather}
where
\begin{equation}
\begin{gathered}
  \beta\left(\alpha\right) =\nu^2\frac{\mathrm{d}}{\mathrm{d}\nu^2}\left(\frac{\alpha}{\pi}\right)=
  -\beta_0\left(\frac{\alpha}{\pi}\right)^2-\beta_1\left(\frac{\alpha}{\pi}\right)^3+\ldots
\\
  \beta_0 = \frac{11}{4}-\frac{1}{6} n_f\quad ,\quad
  \beta_1=\frac{51}{8}-\frac{19}{24}n_f
\end{gathered}
\label{betaDerek}
\end{equation}
or through the correlation function of $I=0,1$ (non-strange) quark currents
\begin{gather}
  \Pi_q\left(Q^2\right) = \ii\,\int\,\mathrm{d}^4x\;\e^{\ii q\cdot x}
  \left\langle 0 | T \left[J_q (x) J_q (0)\right] |0\right\rangle
  \quad,\quad Q^2\equiv -q^2
  \label{quark_corr}
  \\
  J_q(x) = m_q\left[\overline{u}(x)u(x) + (-1)^I\, \overline{d}(x)d(x)\right]/{2}
  \quad.
  \label{quark_curr}
\end{gather}
The quark mass factor  $m_q = \left(m_u + m_d\right)/2$ in \eqref{quark_curr} results in a renormalization-group invariant
current. Although both the gluonic and  ($I=0$) quark correlation functions are probes of scalar hadronic states, those which have a
more significant overlap with the gluonic current will predominate in \eqref{glue_corr}, while those states which are dominantly of a
quark nature are more significant in \eqref{quark_corr}.  A mixed state with substantial gluonic and quark components ({\it i.e.}\ a state that
overlaps with both the gluonic and quark currents) should self-consistently appear in an analysis of both correlation functions.  In
particular,  prediction of mass-degenerate states from the  QCD sum-rule analysis associated with these two correlation functions is
evidence for  interpretation as a mixed state.

In the scalar gluonic channel, the low-energy theorem (LET) \cite{let}
\begin{equation}
  \Pi_g(0)\equiv\lim_{Q^2\rightarrow 0} \Pi_g(Q^2) = \frac{8\pi}{\beta_0} \langle J_g\rangle
\label{letDerek}
\end{equation}
allows construction of the $k=-1$ GSR.  The significance of instanton contributions in the overall consistency of the LET-sensitive
$k=-1$ sum-rule and the LET-insensitive $k\ge 0$ sum-rules was first  demonstrated for Laplace sum-rules \cite{shuryak,gluelet}.  A
similar consistency is observed for the Gaussian sum-rules, but theoretical uncertainties are better controlled in the $k\ge 0$ GSR
\cite{har01}, and hence this paper will focus on the $k=0$ GSRs for both the quark and gluonic channels.

The QCD correlation functions~\eqref{glue_corr} and~\eqref{quark_corr} contain
perturbative, condensate, and instanton contributions.
For the scalar gluonic case, substituting~\eqref{glue_corr} into~\eqref{finishDerek}
and performing the relevant integrals
yields the QCD prediction 
$G_0^{(g)}\left(\hat s,\tau,s_0\right)$
for the $k=0$ GSR to leading order in the quark mass
\begin{equation}
\begin{split}
   G_0^{(g)}(\hat{s},\tau,s_0) =& - \frac{1}{\sqrt{4\pi\tau}} \int\limits_0^{s_0} t^2
   \exp\left[\frac{-(\hat{s}-t)^2}{4\tau}\right]
\Biggl[ (a_0-\pi^2 a_2) + 2a_1\log\left(\frac{t}{\nu^2}\right)
   + 3a_2 \log^2\left(\frac{t}{\nu^2}\right)\Biggr]\dif{t}
\\
   &-\frac{1}{\sqrt{4\pi\tau}} b_1\langle J_g\rangle \int\limits_0^{s_0}
   \exp\left[ \frac{-(\hat{s}-t)^2}{4\tau}\right] \dif{t}
   +\frac{1}{\sqrt{4\pi\tau}} \exp\left( \frac{-\hat{s}^2}{4\tau}\right)
   \left[ c_0 \left\langle {\cal O}_6\right\rangle  - \frac{d_0 \hat{s}}{2\tau}
   \left\langle {\cal O}_8\right\rangle \right]
\\
   &-\frac{16\pi^3}{\sqrt{4\pi\tau}} n_c\rho^4
    \int\limits_0^{s_0} t^2  \exp\left[ \frac{-(\hat{s}-t)^2}{4\tau}\right]  J_2\left(\rho\sqrt{t} \right)
    Y_2\left(\rho\sqrt{t} \right)  \dif{t}
\quad.
\end{split}
\label{G0}
\end{equation}
The perturbative coefficients in~\eqref{G0} are given by
\begin{equation}
\begin{gathered}
  a_0 = -2\left(\frac{\alpha}{\ppi}\right)^2\left[1+\frac{659}{36}\frac{\alpha}{\ppi}+
  247.480\left( \frac{\alpha}{\ppi}\right)^2\right] \\
  a_1 = 2\left(\frac{\alpha}{\ppi}\right)^3\left[ \frac{9}{4}
            +65.781\frac{\alpha}{\ppi}\right] \quad ,\quad
  a_2 = -10.1250\left(\frac{\alpha}{\ppi}\right)^4
\end{gathered}
\label{pertDerek}
\end{equation}
as obtained from the three-loop $\overline{\rm{MS}}$ calculation of the
correlation function in the chiral limit of $n_f=3$ flavours~\cite{che98}.
The condensate contributions in~\eqref{G0} involve next-to-leading order~\cite{bag90}
contributions\footnote{The calculation of next-to-leading contributions in \cite{bag90} have been extended non-trivially to
$n_f=3$ from $n_f=0$, and the operator basis has been changed from $\left\langle \alpha G^2\right\rangle$ to $\langle J_g\rangle$.}
from the dimension four gluon condensate $\langle J_g\rangle$ and leading order~\cite{NSVZ_glue}
contributions from gluonic condensates of dimension six and eight
\begin{gather}
  \left\langle {\cal O}_6\right\rangle  =
  \left\langle g f_{abc}G^a_{\mu\nu}G^b_{\nu\rho}G^c_{\rho\mu}\right\rangle
  \label{dimsix}
\\
  \left\langle {\cal O}_8\right\rangle = 14\left\langle\left(\alpha f_{abc}G^a_{\mu\rho}
  G^b_{\nu\rho}\right)^2\right\rangle
  -\left\langle\left(\alpha f_{abc}G^a_{\mu\nu}G^b_{\rho\lambda}\right)^2\right\rangle
  \label{dimeight}
\\
  \begin{gathered}
    b_0 = 4\ppi\frac{\alpha}{\ppi}\left[ 1+ \frac{175}{36}\frac{\alpha}{\ppi}\right]
    \quad,\quad
    b_1 = -9\ppi\left(\frac{\alpha}{\ppi}\right)^2
    \quad,
  \\
    c_0 = 8\ppi^2\left(\frac{\alpha}{\ppi}\right)^2
    \quad,\quad
    d_0 = 8\ppi^2\frac{\alpha}{\ppi}
    \quad.
  \end{gathered}
  \label{cond_coefficients}
\end{gather}
The remaining term in the GSR~\eqref{G0} represents instanton contributions obtained from
single instanton and anti-instanton~\cite{basic_instanton}
({\it i.e.}\ assuming  that multi-instanton effects are negligible~\cite{schaefer_shuryak})
contributions to the scalar gluonic correlator~\cite{shuryak,gluelet,NSVZ_glue,inst_K2}
within the liquid instanton model~\cite{DIL}
parameterized by the instanton size $\rho$  and  the instanton density $n_c$.
The quantities $J_2$ and $Y_2$ are Bessel functions in the notation of~\cite{abr}.

As a result of renormalization-group scaling of the GSRs~\cite{orl00,gauss}, the coupling in the
perturbative and condensate coefficients (\eqref{pertDerek} and~\eqref{cond_coefficients})
is implicitly the running coupling at the scale $\nu^2=\sqrt{\tau}$
for $n_f=3$ in the $\overline{\rm{MS}}$ scheme
\begin{equation}
\begin{gathered}
  \frac{\alpha (\nu^2)}{\pi} = \frac{1}{\beta_0 L}-\frac{\bar\beta_1\log L}{\left(\beta_0L\right)^2}+
  \frac{1}{\left(\beta_0 L\right)^3}\left[
  \bar\beta_1^2\left(\log^2 L-\log L -1\right) +\bar\beta_2\right]
\\
  L=\log\left(\frac{\nu^2}{\Lambda^2}\right)\quad ,\quad \bar\beta_i=\frac{\beta_i}{\beta_0}
  \quad ,\quad
  \beta_0=\frac{9}{4}\quad ,\quad \beta_1=4\quad ,\quad \beta_2=\frac{3863}{384}
\end{gathered}
\label{run_coupling}
\end{equation}
with $\Lambda_{\overline{MS}}\approx 300\,{\rm MeV}$ for three active flavours,
consistent with current estimates of $\alpha(M_\tau)$~\cite{pdg}.

Analogous to the scalar gluonic case, substitution of~\eqref{quark_corr} into~\eqref{finishDerek}
provides the QCD prediction of $G_0^{(q)}\left(\hat s,\tau,s_0\right)$
for the $I=0,1$ scalar quark currents. To leading order in the quark mass, we find~\cite{orl00}
\begin{equation}
\begin{split}
  G_0^{(q)}\left(\hat s, \tau, s_0\right)=&
  \frac{1}{\sqrt{4\pi\tau}}\frac{3m_q^2}{16\pi^2}
  \int\limits_0^{s_0}
  \exp\left[\frac{-\left(t-\hat s\right)^2}{4\tau}\right]
  \left[t\left(1+\frac{17}{3}\frac{\alpha}{\pi}\right)
  -2\frac{\alpha}{\pi}t\log{\left(\frac{t}{\nu^2}\right)}
  \right]\;\dif{t}
\\
  &+m_q^2\exp{\left(\frac{-\hat s^2}{4\tau}\right)}\left[
  \frac{1}{2\sqrt{\pi\tau}}\left\langle C^s_4{\cal O}^s_4\right\rangle-
  \frac{\hat s}{4\tau\sqrt{\pi\tau}}\left\langle C^s_6{\cal O}^s_6\right\rangle
  \right]
\\
  &-\left(-1\right)^I\frac{3 m_q^2}{8\pi}
  \frac{1}{\sqrt{4\pi\tau}}\int\limits_0^{s_0}
  t\exp\left[\frac{-\left(t-\hat s\right)^2}{4\tau}\right]
  J_1\left(\rho\sqrt{t}\right) Y_1\left(\rho\sqrt{t}\right)\;\dif{t}
  \quad.
\end{split}
\label{gauss_scalar_QCD}
\end{equation}
Again, renormalization-group improvement implies that both
$m_q$ and $\alpha$ are implicitly running quantities at the scale $\nu^2=\sqrt{\tau}$ as given by
\eqref{run_coupling} and the (two-loop, $n_f=3$, $\overline{\rm MS}$) expression
\begin{equation}
m_q\left(\nu^2\right)=  \frac{\hat m_q}{\left(\frac{1}{2}L\right)^{\frac{4}{9}}}\left(
1+\frac{290}{729}\frac{1}{L}-\frac{256}{729}\frac{\log{ L}}{L}
\right) \quad  ,\quad L=\log\left(\frac{\nu^2}{\Lambda^2}\right)\quad ,
\label{run_mass}
\end{equation}
where $\hat m_q$ is the renormalization-group invariant quark mass parameter. The perturbative contributions
in \eqref{gauss_scalar_QCD} are the  $n_f=3$ two-loop results obtained from  \cite{chetyrkin},
and the  instanton expressions are obtained from \cite{SVZ}.     The condensate
contributions are leading-order results obtained from \cite{SVZ}, and are defined by the quantities
\begin{equation}
  \left\langle C_4^s {\cal O}_4^s\right\rangle =
  \frac{3}{2} \left\langle m_q \overline{q}q\right\rangle +
  \frac{1}{16\pi} \left\langle\alpha_s G^2 \right\rangle
  \label{c4_scalar}
\end{equation}
and
\begin{equation}
  \langle C^s_6{\cal O}^s_6\rangle = \pi\alpha_s
  \Biggl[
  \frac{1}{4}\left\langle\left(\bar u\sigma_{\mu\nu}\lambda^a u-\bar
  d\sigma_{\mu\nu}\lambda^a d\right)^2\right\rangle
+\frac{1}{6}
  \left\langle \left(
  \bar u \gamma_\mu \lambda^a u+\bar d \gamma_\mu \lambda^a d
  \right)
  \sum_{u,d,s}\bar q \gamma^\mu \lambda^a q
  \right\rangle\Biggr]\quad .
\label{o6}
\end{equation}
The  vacuum saturation hypothesis \cite{SVZ} in the $SU(2)$ limit
$\langle \bar u u\rangle=\langle\bar d d\rangle\equiv\langle\bar q q\rangle$ provides  a reference value
for $\langle {\cal O}^s_6\rangle$
\begin{equation}
\left\langle C_6^s{\cal O}^s_6\right\rangle=-f_{vs}\frac{88}{27}\alpha_s
\left\langle (\bar q q)^2\right\rangle
=-f_{vs}1.8\times 10^{-4} {\rm GeV}^6\quad ,
\label{o61}
\end{equation}
where the quantity $f_{vs}$ parameterizes deviations from vacuum saturation where $f_{vs}=1$.

The GSRs \eqref{G0} and \eqref{gauss_scalar_QCD} exhibit some interesting qualitative features. For example, the condensate
contributions decay exponentially with the Gaussian peak-position $\hat s$, emphasizing that these contributions have a low-energy
origin. Also, the explicit factor of $I$ appearing in the instanton contributions in the quark scalar channel are a non-perturbative
source of isospin symmetry-breaking.

Before proceeding with an analysis of the GSRs, the QCD input parameters must be specified.
We assume  that
$\langle J_g\rangle\approx\langle\alpha G^2 \rangle$
and then employ the (central) value from \cite{nar97}
\begin{equation}\label{dimfour}
   \langle\alpha G^a_{\ \mu\nu}G^{a\mu\nu}\rangle=\langle\alpha G^2\rangle
      = (0.07\pm 0.01)\, {\rm GeV^4} \quad.
\end{equation}
The dimension six gluon condensate~\eqref{dimsix} can be related to
this value of $\langle\alpha G^2\rangle$ using
instanton techniques (see~\cite{NSVZ_glue,SVZ})
\begin{equation}
   \langle \mathcal{O}_6 \rangle = (0.27\, {\rm GeV}^2)
   \langle \alpha G^2\rangle\quad.
\end{equation}
Further, by invoking vacuum saturation in conjunction with the heavy quark
expansion, the authors of~\cite{bag85} have also related the dimension eight
gluon condensate~\eqref{dimeight} to $\langle\alpha G^2\rangle$
through
\begin{equation}
   \langle\mathcal{O}_8\rangle = \frac{9}{16}
   \left(\langle\alpha G^2\rangle\right)^2\quad.
\end{equation}
In addition, the dilute instanton liquid (DIL) model~\cite{DIL} parameters
\begin{equation} \label{DILparams}
  n_{{c}} = 8.0\times 10^{-4}\ {\rm GeV^4}\quad,\quad\ \rho =
  \frac{1}{0.6}\ {\rm GeV}^{-1}
\end{equation}
will be employed.  Finally, we use $f_{vs}=1.5$ as a central value to accommodate the observed deviations from
vacuum saturation in the dimension six quark condensates~\eqref{o61}~\cite{bordes}.
Note that knowledge of the quark mass parameter is not needed for an analysis based on
the normalized GSRs since $m_q$ provides a common prefactor in~\eqref{gauss_scalar_QCD}.

\section{Analysis of the Gaussian Sum-Rules for Scalar Gluonic and Quark Currents}
In the single narrow resonance  model
\begin{equation}
\rho^{\rm had}(t)=f^2\delta\left(t-m^2\right)
\end{equation}
the  $k=0$ NGSR \eqref{ngsr} becomes
\begin{equation}\label{phenom_single}
  N_0^{\qcd}\left(\hat{s},\tau,s_0\right) = \frac{1}{\sqrt{4\ppi\tau}}
  \exp\left[\frac{-(\hat{s}-m^2)^2}{4\tau}\right]
  \quad.
\end{equation}
The prediction of the mass $m$ is obtained by optimizing the parameter $s_0$ so that the
left-hand side of~\eqref{phenom_single} has
a maximum as a function of $\hat s$ ($\hat s$ peak position) independent of $\tau$ as required by the properties of the right-hand side of~\eqref{phenom_single}~\cite{orl00}.  The  $\tau$-stable $\hat s$ peak  for this optimized $s_0$ then provides the prediction of the resonance mass $m$.

\begin{figure}[htb]
\centering
\includegraphics[scale=0.4]{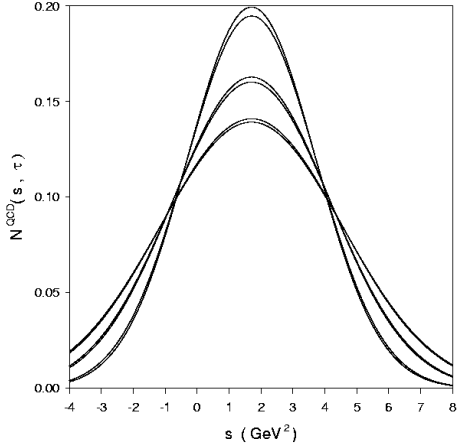}
\caption{
{Comparison of the
theoretical prediction for the  normalized GSR $N_0^{(g)}\left(\hat s, \tau,s_0\right)$ with the
single narrow resonance phenomenological model for the optimized value of the continuum $s_0$.
The $\tau$ values used for the three pairs of curves, from top to bottom in the figure, are respectively
 $\tau=2.0\, {\rm GeV}^4$, $\tau=3.0\,{\rm GeV}^4$, and
$\tau=4.0\,{\rm GeV}^4$. A qualitatively similar agreement between the
 single narrow resonance model and the QCD prediction exists for the $I=0,1$ scalar quark channels.
}}
\label{nar_res_fig}
\end{figure}

The integrity of this procedure has been demonstrated for the vector-isovector currents which probe the $\rho$ meson, resulting in a
 predicted $\rho$ mass of $750\,{\rm MeV}$ and  superb agreement between the phenomenological and QCD sides of the NGSR
\cite{orl00}.  However, the same procedure applied to the scalar gluonic and scalar quark channels leads to significant
disagreement between the two sides of the NGSR as illustrated in Figure~\ref{nar_res_fig}~\cite{orl00,har01}.\footnote{The $\tau$-stability
analysis of the $\hat s$ peak is examined in the range $2\,{\rm GeV^4}\le\tau\le 4\,{\rm GeV^4}$ where the perturbative series is
reasonably convergent while maintaining a Gaussian resolution of typical hadronic scales.}  In both the scalar gluonic and
scalar quark sum-rules, the single resonance model is larger than the theoretical contribution at the peak and underestimates the
theoretical contribution in the tails.  Since both the theoretical and phenomenological contributions are normalized, this implies that
the QCD prediction is broader than the phenomenological model.

\begin{figure}[htb]
\centering
\includegraphics[scale=0.4]{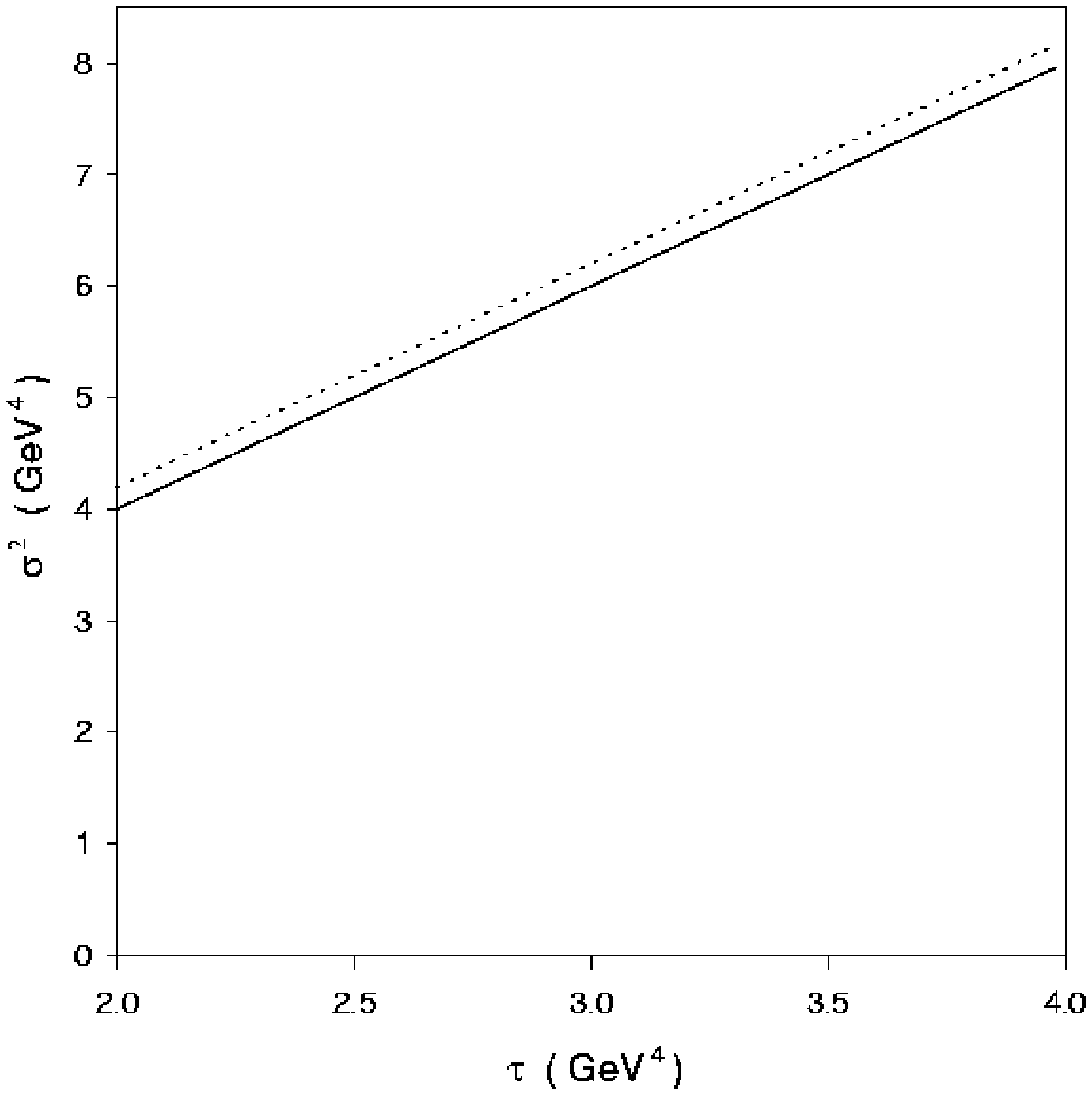}
\caption{Plot of $\sigma^2$ for the theoretical prediction (dotted curve) for the scalar gluonic GSR compared with $\sigma^2=2\tau$
for the single-resonance model (solid curve)
using the optimized value of the continuum. A qualitatively similar result exists for the $I=0,1$ scalar quark channels.}
\label{sigma}
\end{figure}

The following second-order moment combination
\begin{equation}
\label{sigma_combo}
  \sigma^2 \equiv \frac{M_{0,2}}{M_{0,0}} -\left(\frac{M_{0,1}}{M_{0,0}}\right)^2
\end{equation}
provides a quantitative measure of the width of the GSRs.  In the single narrow resonance model we find $\sigma^2=2\tau$, and hence a
significant deviation of the QCD prediction from this result  indicates a failure of the phenomenological model to adequately
describe a particular channel's hadronic content. Figure \ref{sigma} illustrates that $\sigma^2>2\tau$, indicating that a phenomenological model with
distributed resonance strength is necessary \cite{orl00,har01}.

A phenomenological model with two narrow resonances of mass $m_1$ and $m_2$ results in the NGSR
\begin{equation}
N_0\left(\hat s, \tau,s_0\right)=\frac{r_1}{\sqrt{4\pi\tau}}
\exp{\left[
\frac{-\left(\hat s-m_1^2\right)^2}{4\tau}\right]}
+\frac{r_2}{\sqrt{4\pi\tau}}\exp{\left[
\frac{-\left(\hat s-m_2^2\right)^2}{4\tau}\right]}
\label{norm_gauss_2res}
\end{equation}
where $r_1+r_2=1$ describes the relative strength of the two resonances contributing to the spectral function.
In terms of the parameters $\{ r,y,z\}$ defined by
\begin{equation}
r=r_1-r_2\,,~y=m_1^2-m_2^2\,,~z=m_1^2+m_2^2\quad,
\end{equation}
the second-order moment combination~\eqref{sigma_combo} resulting from the left-hand side of~\eqref{norm_gauss_2res}
\begin{equation}
\sigma^2-2\tau=\frac{1}{4}y^2\left(1-r^2\right)>0
\end{equation}
naturally results in a broader distribution than the single resonance model as suggested by the QCD result.

Analysis of the double narrow resonance model is substantially more complicated than the single narrow resonance case.  In particular,
the $\hat s$ peak position (denoted by $\hat s_{peak}$) develops $\tau$ dependence which is well-described by \cite{orl00}
\begin{equation}
\hat s_{peak}\left(\tau,s_0\right)=  A + \frac{B}{\tau} + \frac{C}{\tau^2}
\end{equation}
and hence $s_0$ is optimized to obtain the best description of this $\tau$ dependence. After optimization of $s_0$, the parameters in
the phenomenological model can be determined from the moment combinations
\begin{eqnarray}
& &  z  =   2\frac{M_{0,1}}{M_{0,0}} + \frac{A_2}{\sigma^2-2\tau}
\label{z_moms}\\
& &  y  =   \frac{ -\sqrt{A_2^2 + 4(\sigma^2-2\tau)^3}}{\sigma^2-2\tau}
\label{y_moms}\\
& &  r  =   \frac{A_2}{\sqrt{A_2^2 + 4(\sigma^2-2\tau)^3}}
\label{r_moms}
\end{eqnarray}
where the third-order moment combination $A_2$, representing the asymmetry of the distribution, is defined by
\begin{equation}
A_2=\frac{M_{0,3}}{M_{0,0}}-3\frac{M_{0,2}}{M_{0,0}}\frac{M_{0,1}}{M_{0,0}}
+2\left(\frac{M_{0,1}}{M_{0,0}}\right)^3 \quad .
\label{dist_asymm}
\end{equation}
This procedure for optimizing $s_0$ and determining the resonance parameters has been confirmed by a more numerically-intensive
multi-parameter fit of $s_0$ and the resonance parameters \cite{orl00,ps_glue}.

The  resonance parameters resulting from this analysis  shown in Table \ref{doubres_tab} \cite{orl00,har01}
illustrate a remarkably consistent scenario of a
$1\,{\rm GeV}$ and a $1.4\,{\rm GeV}$ state coupled to both the gluonic and non-strange $I=0$ quark currents, with the heavier state
more strongly coupled to the gluonic operators.  This consistency of the mass predictions in the two channels is precisely what is
expected for hadronic states which are mixtures of gluonium and quark mesons.
The results in the $I=1$ scalar quark channel support the interpretation of the $a_0(1450)$ as the lightest state with a dominant
coupling to the scalar quark currents.

\begin{table}[htb]
  \centering
  \begin{tabular}{||c|c|c|c|c|c||}
    \hline\hline
Sum-Rule &  $m_1$ (GeV) & $m_2$ (GeV)  & $r_1$ & $r_2$ & $s_0$ (${\rm GeV^2}$)  \\
    \hline\hline
gluonic   &  $0.98$ & $1.4$ & $0.28$ & $0.72$ & 2.3 \\\hline
quark $I=0$  & $0.97$ & $1.4$ & $0.63$ & $0.37$ & 2.6\\\hline
quark $I=1$ & $1.4$ & $1.8$ & $0.57$ & $0.43$ & 3.9\\
    \hline\hline
  \end{tabular}
 \caption{Analysis results from scalar quark and gluonic  Gaussian sum-rules in the double narrow resonance model.}
\label{doubres_tab}
\end{table}

\begin{figure}[htb]
\centering
\includegraphics[scale=0.4]{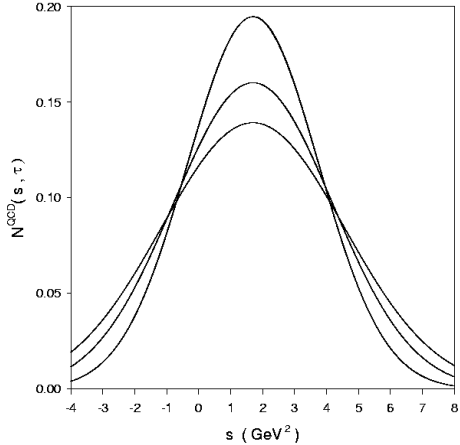}
\caption{
Comparison of the
theoretical prediction  $N_0^{(g)}\left(\hat s, \tau,s_0\right)$ with the
double narrow resonance phenomenological model using the parameters in Table \protect\ref{doubres_tab}.
The $\tau$ values used for the three pairs of curves, from top to bottom in the figure, are respectively
 $\tau=2.0\,{\rm GeV}^4$, $\tau=3.0\, {\rm GeV}^4$, and $\tau=4.0\,
{\rm GeV}^4$.
Note the almost perfect overlap between the theoretical prediction and the
phenomenological models.
A qualitatively similar agreement between the
 double narrow resonance model and the QCD prediction exists for the $I=0,1$ scalar quark channels.
}
\label{twores_fig}
\end{figure}

The double narrow resonance model results in excellent agreement with QCD as illustrated in Figure \ref{twores_fig}.
More complicated resonance models which extend the double narrow resonance model to introduce resonance widths do not improve the quantitative agreement with QCD exhibited in the figure.

In summary, analysis of the GSRs for the scalar gluonic and $I=0$ non-strange quark
scalar currents exhibit  a remarkable similarity in their mass predictions within a
double narrow resonance model, a result indicative of the existence of hadronic states which are
mixtures of quark mesons and gluonium.
Although the effect of QCD uncertainties arising from a 15\% variation in the DIL parameters~\eqref{DILparams} and the $d=4$ gluon condensate~\eqref{dimfour}, as well as variations of the vacuum saturation parameter within the range $1<f_{vs}<2$ correspond to an uncertainty in the Table~\ref{doubres_tab} mass parameters of approximately $0.2\,{\rm GeV}$, we note that the mass splitting of $0.4\,{\rm GeV}$ between the states is remarkably stable \cite{har01}.
This QCD evidence for mixed states with a mass splitting of $0.4\,{\rm GeV}$ provides valuable information for interpretation of the known $f_0$ resonances.

\section{Acknowledgements}
TGS is grateful for research support from the Natural Sciences \&
Engineering Research Council of Canada (NSERC).
DH is thankful for support from the Department of Research at the
University College of the Fraser Valley (UCFV).
Many thanks to Amir Fariborz
for his efforts in organizing the Utica Workshop on Scalar Mesons
which resulted in a tremendously  enjoyable and valuable workshop.

\end{document}